\documentclass[a4paper,oneside]{amsart}

\RequirePackage{amsmath}
\RequirePackage{bm}
\RequirePackage{amssymb}
\RequirePackage{upref}
\RequirePackage{amsthm}
\RequirePackage{enumerate}
\RequirePackage{pb-diagram}
\RequirePackage{amsfonts}
\RequirePackage[mathscr]{eucal}
\RequirePackage{verbatim}
\RequirePackage{xr}
\RequirePackage{graphicx}
\RequirePackage[pdftex]{floatflt}
\usepackage{calc}
\usepackage{xspace}

\newcommand{\cf}{cf.\@\xspace}

\newcommand{\al}{\alpha}
\newcommand{\bet}{\beta}
\newcommand{\ga}{\gamma}
\newcommand{\de}{\delta }

\newcommand{\f}{\varphi}
\newcommand{\h}{\eta}

\newcommand{\ka}{\kappa}
\newcommand{\lam}{\lambda}

\newcommand{\om}{\omega}

\newcommand{\s}{\sigma}
\newcommand{\x}{\xi}

\newcommand{\Lam}{\varLambda}

\newcommand{\so}{{\mc S_0}}

\newcommand{\const}{\tup{const}}
\newcommand{\ndash}{\nobreakdash--}

\newcommand{\msp[1]}[1]{\mspace{#1mu}}

\newcommand{\R}[1][n+1]{{\protect\mathbb R}^{#1}}

\newcommand{\N}{{\protect\mathbb N}}

\newcommand{\eR}{\stackrel{\lower1ex \hbox{\rule{6.5pt}{0.5pt}}}{\msp[3]\R[]}}
\newcommand{\eN}{\stackrel{\lower1ex \hbox{\rule{6.5pt}{0.5pt}}}{\msp[1]\N}}
\newcommand{\eO}{\stackrel{\lower1ex
\hbox{\rule{6pt}{0.5pt}}}{\msc O}}

\DeclareMathOperator{\graph}{graph}

\newcommand\ra{\rightarrow}

\newcommand{\ua}{\uparrow}

\newcommand{\un}{\infty}
\newcommand{\A}{\forall}

\newcommand{\set}[2]{\{\,#1\colon #2\,\}}
\newcommand{\uu}{\cup}
\newcommand{\ii}{\cap}
\newcommand{\uuu}{\bigcup}

\newcommand{\uud}{ \stackrel{\lower 1ex \hbox {.}}{\uu}}
\newcommand{\uuud}[1]{ \stackrel{\lower 1ex \hbox {.}}{\uuu_{#1}}}

\newcommand{\sminus}[1][28]{\raise 0.#1ex\hbox{$\scriptstyle\setminus$}}

\newcommand\ti{\times }

\newcommand{\abs}[1]{\lvert#1\rvert}

\newcommand{\tup}{\textup}

\newcommand{\mc}{\protect\mathcal}
\newcommand{\msc}{\protect\mathscr}

\providecommand{\bysame}{\makebox[3em]{\hrulefill}\thinspace}

\newcommand{\bt}{\begin{thm}}
\newcommand{\bl}{\begin{lem}}
\newcommand{\bc}{\begin{cor}}
\newcommand{\bd}{\begin{definition}}
\newcommand{\bpp}{\begin{prop}}
\newcommand{\br}{\begin{rem}}
\newcommand{\bn}{\begin{note}}
\newcommand{\be}{\begin{ex}}
\newcommand{\bes}{\begin{exs}}
\newcommand{\bb}{\begin{example}}
\newcommand{\bbs}{\begin{examples}}
\newcommand{\ba}{\begin{axiom}}

\newcommand{\et}{\end{thm}}
\newcommand{\el}{\end{lem}}
\newcommand{\ec}{\end{cor}}
\newcommand{\ed}{\end{definition}}
\newcommand{\epp}{\end{prop}}
\newcommand{\er}{\end{rem}}
\newcommand{\en}{\end{note}}
\newcommand{\ee}{\end{ex}}
\newcommand{\ees}{\end{exs}}
\newcommand{\eb}{\end{example}}
\newcommand{\ebs}{\end{examples}}
\newcommand{\ea}{\end{axiom}}

\newcommand{\bp}{\begin{proof}}
\newcommand{\ep}{\end{proof}}
\newcommand{\eps}{\renewcommand{\qed}{}\end{proof}}

\newcommand{\bal}{\begin{align}}

\newcommand{\bi}[1][1.]{\begin{enumerate}[\upshape #1]}
\newcommand{\bia}[1][(1)]{\begin{enumerate}[\upshape #1]}
\newcommand{\bin}[1][1]{\begin{enumerate}[\upshape\bfseries #1]}
\newcommand{\bir}[1][(i)]{\begin{enumerate}[\upshape #1]}
\newcommand{\bic}[1][(i)]{\begin{enumerate}[\upshape\hspace{2\cma}#1]}
\newcommand{\bis}[2][1.]{\begin{enumerate}[\upshape\hspace{#2\parindent}#1]}
\newcommand{\ei}{\end{enumerate}}

\newcommand\ndots{\raise 0.47ex \hbox {,}\hskip0.06em\cdots %
     \raise 0.47ex \hbox {,}\hskip0.06em} 

\newcommand{\q}{\quad}
\newcommand{\qq}{\qquad}

\newcommand{\hp}{\hphantom}

\newcommand\nd{\noindent}

\newskip\Csmallskipamount                                                
\Csmallskipamount=\smallskipamount
\newskip\Cmedskipamount
\Cmedskipamount=\medskipamount
\newskip\Cbigskipamount
\Cbigskipamount=\bigskipamount

\newcommand\cvs{\vspace\Csmallskipamount}   
\newcommand\cvm{\vspace\Cmedskipamount}

\newskip\csa
\csa=\smallskipamount

\newskip\cma
\cma=\medskipamount

\newskip\cba
\cba=\bigskipamount

\newdimen\spt
\newcommand\citem{\cvs\advance\itemno by
1{(\romannumeral\the\itemno})\hskip3pt}
\newcommand{\bitem}{\cvm\nd\advance\itemno by
1{\bf\the\itemno}\hspace{\cma}}

\newcount\itemno
\newcommand{\las}[1]{\label{S:#1}}

\newcommand{\lae}[1]{\label{E:#1}}
\newcommand{\lat}[1]{\label{T:#1}}
\newcommand{\lal}[1]{\label{L:#1}}

\newcommand{\rs}[1]{Section~\ref{S:#1}}

\newcommand{\rl}[1]{Lemma~\ref{L:#1}}

\newcommand{\re}[1]{\eqref{E:#1}}

\newskip\thmskip
\thmskip=\parindent

\newskip\hsk
\newenvironment{hinw}{\labelsep=0pt\begin{list}{}{\labelsep=0pt\itemindent=0pt\labelwidth=0pt\leftmargin=\parindent\rightmargin=0pt\partopsep=\cba}%
\item\it\nopagebreak\nopagebreak}%
{\end{list}}

\newcommand\bh{\begin{hinw}}
\newcommand{\eh}{\end{hinw}}

\newtheoremstyle{normal}
  {\cba}
  {\cba}
  {}
  {\thmskip}
  {\bfseries}
  {.}
  {\hsk}
  {}

\newtheoremstyle{abschnitt}
  {\cba}
  {\cba}
  {}
  {\thmskip}
  {\bfseries}
  {.}
  {\hsk}
  {}

\newtheoremstyle{italic}
  {\cba}
  {\cba}
  {\itshape}
  {\thmskip}
  {\bfseries}
  {.}
  {\hsk}
  {}

\newtheoremstyle{aufgaben}
  {\cba}
  {\cba}
  {}
  {}
  {\normalsize\bfseries}
  {.}
  {\hsk}
  {}

\newtheoremstyle{break}
  {\cba}
  {\cba}
  {\itshape}
  {}
  {\bfseries}
  {.}
  {\newline}
  {}

\swapnumbers
\theoremstyle{italic}
\newtheorem{thm}[subsection]{Theorem}
\newtheorem{lem}[subsection]{Lemma}
\newtheorem{prop}[subsection]{Proposition}
\newtheorem{cor}[subsection]{Corollary}

\theoremstyle{normal}
\newtheorem{rem}[subsection]{Remark}
\newtheorem{definition}[subsection]{Definition}
\newtheorem{example}[subsection]{Example}
\newtheorem{examples}[subsection]{Examples}
\newtheorem{ex}[subsection]{Exercise}
\newtheorem{note}[subsection]{}
\newtheorem{axiom}[subsection]{Axiom}

\theoremstyle{aufgaben}
\newtheorem{exs}[subsection]{Exercises}

\swapnumbers

\numberwithin{equation}{section}
\numberwithin{figure}{section}

\newenvironment{textequation}[1][0.8]
{\begin{equation}
\begin{aligned}
\begin{minipage}{#1\linewidth}}
{\end{minipage}
\end{aligned}
\end{equation}
\ignorespacesafterend}

\newcommand{\btext}{\begin{textequation}}
\newcommand{\etext}{\end{textequation}}

\usepackage[german,english]{babel}
\usepackage{graphicx}
\RequirePackage{amsmath}
\RequirePackage{bm}
\RequirePackage{amssymb}
\RequirePackage{upref}
\RequirePackage{amsthm}
\RequirePackage{enumerate}
\RequirePackage{pb-diagram}
\RequirePackage{amsfonts}
\RequirePackage[mathscr]{eucal}
\RequirePackage{verbatim}
\RequirePackage{xr}
\RequirePackage{graphicx}
\RequirePackage[pdftex]{floatflt}
\usepackage{calc}
\usepackage{xspace}

\RequirePackage{upref}
\RequirePackage{amsthm}
\RequirePackage{enumerate}
\usepackage[mathscr]{eucal}

\usepackage{xr-hyper}

\usepackage{calc}

\newlength{\oddsidemarginlength}
\newlength{\topmarginlength}

\newcounter{numberoflines}
\newcounter{tempcc}
\oddsidemargin=\oddsidemarginlength
\evensidemargin=\oddsidemargin
\topmargin=\topmarginlength
\usepackage{hyperref}

\begin{document}

\flushbottom


\title[The inverse mean curvature flow in Robertson-Walker spaces]{The inverse mean
curvature flow in Robertson-Walker spaces and its application to cosmology}

\author{Claus Gerhardt}
\address{Ruprecht-Karls-Universit\"at, Institut f\"ur Angewandte Mathematik,
Im Neuenheimer Feld 294, 69120 Heidelberg, Germany}
\email{gerhardt@math.uni-heidelberg.de}
\urladdr{http://www.math.uni-heidelberg.de/studinfo/gerhardt/}
\thanks{}

%
\date{\today}
%


\begin{abstract}
We consider the inverse mean curvature flow in Robertson-Walker spacetimes that
satisfy the Einstein equations and have a big crunch singularity and prove that under
natural conditions the rescaled inverse mean curvature flow provides a smooth
transition from big crunch to big bang. We also construct an example showing that in
general the transition flow is only of class $C^3$.
\end{abstract}

\maketitle

\tableofcontents

\setcounter{section}{-1}
\section{Introduction}

In a recent paper \cite{cg:arw} we proved that the inverse mean curvature flow,
properly rescaled, did provide a transition from big crunch to big bang. The
transition flow was of class $C^3$. The underlying $(n+1)$-dimensional spacetime
$N$ was fairly general, a cosmological spacetime satisfying some structural
conditions, we called these spacetimes ARW spaces.

In this paper we shall show that in general the differentiability class $C^3$ is the best
posible for the transition flow. If it should be of class $C^\un$, then additional
assumptions have to be satisfied.

\cvm
We shall consider the problem in Robertson-Walker spaces $N=I\ti \so$, where $\so$
is a spaceform with curvature $\tilde\ka=-1,0,1$, it may be compact or not, and the
metric in $N$ is of the form
\begin{equation}
d\bar s^2=e^{2f}(-(dx^0)^2+\s_{ij}(x)dx^idx^j),
\end{equation}
where $x^0=\tau$ is the time function, $(\s_{ij})$ the metric of $\so$, 
$f=f(\tau)$, and $x^0$ ranges between $-a<x^0<0$. We assume that there is a big
crunch singularity in $\{x^0=0\}$, i.e., we assume
\begin{equation}\lae{0.2}
\lim_{\tau\ra 0}f(\tau)=-\un\q\tup{and}\q \lim_{\tau\ra 0}-f'=\un.
\end{equation}

The Einstein equations should be valid with a cosmological constant $\Lam$
\begin{equation}
G_{\al\bet}+\Lam\bar g_{\al\bet}=\ka T_{\al\bet},\q\ka>0,
\end{equation}
or equivalently,
\begin{equation}\lae{0.4}
G_{\al\bet}=\ka(T_{\al\bet}-\s\bar g_{\al\bet}),\q \s=\tfrac\Lam\ka.
\end{equation}

If $(T_{\al\bet})$ is the stress-energy tensor of a perfect fluid
\begin{equation}
T^0_0=-\rho,\q T^\al_i=p\de^\al_i
\end{equation}
with an equation of state
\begin{equation}\lae{0.6}
p=\tfrac\om{n}\rho,
\end{equation}
then the equation \re{0.4} is equivalent to the Friedmann equation
\begin{equation}\lae{0.7}
\abs{f'}^2=-\tilde\ka +\tfrac{2\ka}{n(n-1)}(\rho+\s)e^{2f},
\end{equation}
which can be easily derived by looking at the component $\al=\bet=0$ in \re{0.4}.

\cvm
Assuming that $\om$ is of the form
\begin{equation}
\om=\om_0+\lam(f),\q \om_0=\const,
\end{equation}
where $\lam=\lam(t)$ is smooth satisfying
\begin{equation}\lae{0.9}
\lim_{t\ra-\un}\lam(t)=0,
\end{equation}
such that there exists a primitive $\tilde\mu=\tilde\mu(t)$, $\tilde\mu'=\lam$, with
\begin{equation}\lae{0.10}
\lim_{t\ra-\un}\tilde\mu(t)=0,
\end{equation}
then $\rho$ obeys the conservation law
\begin{equation}
\rho=\rho_0e^{-(n+\om_0)f}e^{-\tilde\mu},
\end{equation}
\cf \cite[Lemma 0.2]{cg:arw}. Hence we deduce from \re{0.7}
\begin{equation}\lae{0.12}
\abs{f'}^2=-\tilde\ka
+\tfrac{2\ka}{n(n-1)}(\rho_0e^{-(n+\om_0)f}e^{-\tilde\mu}+\s)e^{2f}.
\end{equation}

The main result of the paper can be summarized in the following theorem

\bt
Let $\tilde\ga=\tfrac12(n+\om_0-2)>0$, and assume that $\lam$ satisfies the
condition \re{1.2} and that
$\mu$ can be viewed as a smooth and even function in the variable
$(-r)^{\tilde\ga}$, where
$r=-e^f<0$, or that it can be extended to a smooth and even function on
$(-\tilde\ga^{-1},\tilde\ga^{-1})$, then the transition flow $y=y(s,\xi)$, as defined
in \re{2.16} and \re{2.17},  is smooth in
$(-\tilde\ga^{-1},\tilde\ga^{-1})\ti\so$, if either
\begin{equation}
\om_0\in\R[]\q\tup{and}\q \s=0,
\end{equation}
or
\begin{equation}
\om_0=4-n\q\tup{and}\q\s\in\R[].
\end{equation}
\et

Let us emphasize that the smooth transition from big crunch to big bang does not
constitute the existence of a cyclic universe, \cf the end of \rs{2} for a detailed
discussion.

In \rs{3} we prove that in general the transition flow is only of class $C^3$ by
constructing a counter example.

\cvm
We believe that the results and even more the proofs indicate strongly that the inverse
mean curvature flow is the right vehicle to offer a smooth transition from big crunch
to big  bang in case of abstract spacetimes that are not embedded in a bulk
spacetime.

We refer to \cite[Section 2]{cg:indiana} for a description of our notations and
conventions.

\section{The Friedmann equation}

We want to solve the Friedmann equation \re{0.12} in an interval $I=(-a,0)$ such
that the resulting spacetime $N$ is an ARW space, \cf \cite[Definition 0.8]{cg:brane}
for a definition that applies to closed spacetimes as well as to non-closed.

In a slightly different setting we proved in \cite[Section 9]{cg:arw} that a
cosmological spacetime satisfying the Einstein equations for a perfect fluid with an
equation of state \re{0.6}, $\om=\const$, is an ARW space, if
\begin{equation}
\tilde\ga=\tfrac12(n+\om-2)>0.
\end{equation}

This result will also be valid in the present situation.

\bl\lal{1.1}
Let $\tilde\ga=\tfrac12(n+\om_0-2)$ be positive and assume that $\lam$ and
$\tilde\mu$ satisfy the conditions stated in the previous section,
and in addition suppose
\begin{equation}\lae{1.2}
\abs{D^m\lam(t)}\le c_m\qq\A\; m\in\N.
\end{equation}
 Then the Friedmann
equation \re{0.12} can be solved in an interval $I=(-a,0)$ such that $f\in C^\un(I)$
and the relations \re{0.2} are valid.
 Moreover, $N$ is an ARW
space.
\el

\bp
We want to apply the existence result \cite[Theorem 3.1]{cg:brane}. Multiply
equation \re{0.12} by $e^{2\tilde\ga f}$ and set
\begin{equation}
\f=e^{\tilde\ga f}
\end{equation}
and
\begin{equation}
r=-e^f.
\end{equation}
Then $\f$ satisfies the differential equation
\begin{equation}\lae{1.5}
\tilde\ga^{-2}\dot\f^2=-\tilde\ka e^{2\tilde\ga f}+\tfrac{2\ka}{n(n-1)}
(\s e^{2(\tilde\ga+1) f}+\rho_0e^{-\mu}),
\end{equation}
where we defined $\mu=\mu(r)$ by
\begin{equation}
\mu(r)=\tilde\mu(\log(-r)).
\end{equation}

Suppose the Friedmann equation were solvable with $f$ satisfying \re{0.2}, then the
right-hand side of \re{1.5} would tend to $\tfrac{2\ka}{n(n-1)}\rho_0$, if $\tau\ra
0$. Thus, we see that solving \re{0.12} and \re{0.2} is equivalent to solving
\begin{equation}\lae{1.7}
\tilde\ga^{-1}\dot\f=-\sqrt{F(\f)}
\end{equation}
with initial value $\f(0)=0$, where
\begin{equation}\lae{1.8}
F(\f)=-\tilde\ka
\f^2+\tfrac{2\ka}{n(n-1)}(\rho_0e^{-\mu}+\s\f^{2(1+\tilde\ga^{-1})})
\end{equation}
and $\mu$ should be considered to depend on
\begin{equation}
\mu(r)=\mu(-\f^{\tilde\ga^{-1}}).
\end{equation}

\cvm
We can now apply the existence result in \cite[Theorem 3.1]{cg:brane} to conclude
that \re{1.7} has a solution $\f\in C^1((-a,0])\ii C^\un((-a,0))$, where, if we choose
$a$ maximal, $a$ is determined by the requirement
\begin{equation}
\lim_{\tau\ra -a}\f=\un\q\tup{or}\q \lim_{\tau\ra -a}F(\f)=0.
\end{equation}

Set $f=\tilde\ga^{-1}\log\f$, then $f$ satisfies \re{0.2}, since $F(0)>0$.

Moreover, differentiating \re{1.5} with respect to $\tau$ and dividing the resulting
equation by $2f' e^{\tilde\ga f}$ we obtain
\begin{equation}
f''+\tilde\ga \abs{f'}^2=-\tilde\ka\tilde\ga +\tfrac\ka{n(n-1)}(2\s (\tilde\ga
+1)e^{2 f}-\lam \rho_0 e^{-\mu}),
\end{equation}
from which we conclude that $N$ is an ARW space, in view of \re{0.9}, \re{0.10} and
\re{1.2}.
\ep

\section{The transition flow}\las{2}

Let $M_0$ be a spacelike hypersurface with positive mean curvature with respect to
the past directed normal, then the inverse mean curvature flow with initial
hypersurface $M_0$ is given by the evolution equation
\begin{equation}\lae{2.1}
\dot x=-H^{-1}\nu,
\end{equation}
where $\nu$ is the past directed normal of the flow hypersurfaces $M(t)$ which are
locally defined by an embedding
\begin{equation}
x=x(t,\xi),\q \x=(\x^i),
\end{equation}
\cf \cite{cg:arw} for details.

In general, even in Robertson-Walker spaces, this evolution problem can only be solved,
if $\so$ is compact. However, if, in the present situation, we assume that $M_0$ is a
coordinate slice $\{x^0=\const\}$, then the fairly complex parabolic system \re{2.1}
is reduced to a scalar ordinary differential equation.

Look at the component $\al=0$ in \re{2.1}. Writing the hypersurfaces $M(t)$ as
graphs over $\so$
\begin{equation}
M(t)=\set{(u,x)}{x\in\so},
\end{equation}
we see that $u$ only depends on $t$, $u=u(t)$, and $u$ satisfies the differential
equation
\begin{equation}\lae{2.4}
\dot u=\frac1{-nf'},
\end{equation}
where $f=f(u)$, with initial value $u(0)=u_0$, \cf \cite[Section 2]{cg:arw}. The
mean curvature of the slices $M(t)$ is given by
\begin{equation}
H=e^{-f}(-nf').
\end{equation}

From \re{2.4} we immediately deduce
\begin{equation}
\tfrac{d}{dt}(nf+t)=nf'\dot u+1=0,
\end{equation}
and hence
\begin{equation}
e^{nf}e^t=\const=c,
\end{equation}
or equivalently,
\begin{equation}\lae{2.8}
e^{\tilde\ga f}e^{\ga t}=c,
\end{equation}
where $\ga =\tfrac1n \tilde\ga$, and where  the
 symbol $c$ may represent different constants.

The conservation law \re{2.8} can be viewed as the integrated version of the inverse
mean curvature flow.

In \cite[Theorem 3.6]{cg:arw} we proved that there are positive constants $c_1,
c_2$ such that
\begin{equation}
-c_1\le \tilde u\le -c_2<0.
\end{equation}
The old proof also works in the present situation, where $\so$ is not necessarily
compact, since $u$ doesn't depend on $x$.

Moreover, 
\begin{equation}
\lim_{t\ra\un}\tilde u\q\tup{exists},
\end{equation}
\cf \cite[Lemma 7.1]{cg:arw}.

We shall define a new spacetime $\hat N$ by reflection and time reversal such that
the IMCF in the old spacetime transforms to an IMCF in the new one.

\cvm
By switching the light cone we obtain a new spacetime $\hat N$. The flow equation
in $N$ is independent of the time orientation, and we can write it as
\begin{equation}
\dot x=- H^{-1}\nu=-(- H)^{-1}(-\nu)\equiv -\hat
H^{-1}\hat \nu,
\end{equation}
where the normal vector $\hat \nu=-\nu$ is past directed in $\hat N$ and the
mean curvature $\hat H=- H$ negative.

Introducing a new time function $\hat x^0=-x^0$ and formally new coordinates
$(\hat x^\al)$ by setting
\begin{equation}
\hat x^0=-x^0,\q\hat x^i=x^i,
\end{equation}
we define a spacetime $\hat N$ having the same metric as $N$---only expressed in
the new coordinate system---such that the flow equation has the form
\begin{equation}\lae{8.3}
\dot{\hat x}=-\hat H^{-1}\hat \nu,
\end{equation}
where $M(t)=\graph \hat u(t)$, $\hat u=-u$. 

\cvm
The singularity in $\hat x^0=0$ is now a past singularity, and can be referred to as a
big bang singularity.

\cvm
The union $N\uu\hat N$ is a smooth manifold, topologically a product
$(-a,a)\ti\so$---we are well aware that formally the singularity $\{0\}\ti\so$ is not
part of the union; equipped with the respective metrics and time orientation it is a
spacetime which has a (metric) singularity in
$x^0=0$. The time function
\begin{equation}\lae{8.7}
\hat x^0=
\begin{cases}
\hp{-}x^0, &\tup{in } N,\\
-x^0, &\tup{in } \hat N,
\end{cases}
\end{equation}
is smooth across the singularity and future directed.

\cvm
Using the time function in \re{8.7} the inverse mean curvature flows in $N$ and
$\hat N$ can be uniformly expressed in the form
\begin{equation}\lae{8.8}
\dot{\hat x}=-\hat H^{-1}\hat\nu,
\end{equation}
where \re{8.8} represents the original flow in $N$, if $\hat x^0<0$, and the flow in
\re{8.3}, if $\hat x^0>0$.

\cvm
In \cite{cg:arw} we then introduced a new flow parameter
\begin{equation}\lae{2.16}
s=
\begin{cases}
-\ga^{-1}e^{-\ga t},& \tup{for the flow in } N,\\
\hp{-}\ga ^{-1}e^{-\ga t},& \tup{for the flow in } \hat N,
\end{cases}
\end{equation}
and defined the flow $y=y(s)$ by $y(s)=\hat x(t)$. $y=y(s)$ is then defined in
$[-\ga^{-1},\ga^{-1}]\times \so$, smooth in $\{s\ne 0\}$, and satisfies the
evolution equation
\begin{equation}\lae{2.17}
y'\equiv \tfrac d{ds}y=
\begin{cases}
-\hat H^{-1}\hat\nu \msp e^{\ga t}, & s<0,\\
\hp{-}\hat H^{-1}\hat\nu \msp e^{\ga t},& s>0,
\end{cases}
\end{equation}
or equivalently, if we only consider the salar version with $\h=\h(s)$ representing
$y^0$
\begin{equation}\lae{2.18}
\h'=\tfrac d{ds}\h=
\begin{cases}
\hp{-}\dot ue^{\ga t},&s<0,\\
-\dot{\hat u}e^{\ga t},& s>0.
\end{cases}
\end{equation}

According to the results in \cite[Theorem 8.1]{cg:arw} $y$, and hence $\h$, are of
class $C^3$ across the singularity.

Now, looking at the relation \re{2.8} we see that the new parameter $s$ could just
as well be defined by
\begin{equation}\lae{2.19}
s=
\begin{cases}
-\tilde\ga^{-1} e^{\tilde\ga f}, & s<0,\\
\hp{-}\tilde\ga^{-1} e^{\tilde\ga f}, & s>0,
\end{cases}
\end{equation}
where in $N$ as well as in $\hat N$ $f$ is considered to be a function of $u(t)$,
$f=f(u(t))$.

Defining $s$ by \re{2.19} we deduce for $s<0$
\begin{equation}\lae{2.20}
\h'=\dot u\tfrac{dt}{ds}=\dot u\frac1{-f'e^{\tilde\ga f}\dot u}=\frac1{-f'
e^{\tilde\ga f}}\equiv \f^{-1}.
\end{equation}
The same relation is also valid for $s>0$. 

\cvm
Suppose now that $\f$, or equivalently, $\f^2$,
\begin{equation}
\f^2=\abs{f'}^2e^{2\tilde\ga f}=-\tilde\ka e^{2\tilde\ga
f}+\tfrac{2\ka}{n(n-1)}(\s e^{2(\tilde\ga +1)f}+\rho_0e^{-\mu}),
\end{equation}
can be viewed as an even function in $e^{\tilde\ga f}$, or equivalently, an even
function in $s$, then $\h$ would be of class $C^\un$ across the singularity, and
hence the transition flow $y=y(s)$ would be smooth.

We have thus proved

\bt\lat{2.1}
Let $\tilde\ga=\tfrac12(n+\om_0-2)>0$, and assume that $\lam$ satisfies the
condition \re{1.2} and that
$\mu$ is a smooth and even function in the variable $(-r)^{\tilde\ga}$, $r<0$, or can
be extended to a smooth and even function on $(-\tilde\ga^{-1},\tilde\ga^{-1})$,
then the transition flow $y=y(s,\xi)$ is smooth in
$(-\tilde\ga^{-1},\tilde\ga^{-1})\ti\so$, if either
\begin{equation}
\om_0\in\R[]\q\tup{and}\q \s=0,
\end{equation}
or
\begin{equation}\lae{2.23}
\om_0=4-n\q\tup{and}\q\s\in\R[].
\end{equation}
\et

If $n=3$ and \re{2.23} is valid, this means that we consider a radiation dominated
universe.

\cvm
Let us also emphasize that in the preceding theorem we have only proved a smooth
transition from big crunch to big bang. This does not necessarily mean that we have a
cyclic universe---the same observation also applies to the transition results we
obtained in \cite{cg:brane} in a brane cosmology setting.

It could well be that the following scenario holds: The spacetime $N$ exists in
$-\un<\tau<0$ with the only singularity in $\tau=0$, a big crunch; the mean
curvature of the slices $\{x^0=\const\}$ is always positive and
\begin{equation}
\lim_{\tau\ra-\un}e^f=\un.
\end{equation}
After a smooth transition through the singularity   the mirror image $\hat N$
develops.

\cvm
Such a pair of universes $(N,\hat N)$ can be easily constructed, in fact, this will
always be the case, if the right-hand side of equation \re{1.8} never vanishes and
grows at most quadratically in $\f$, which will be the case, if $\s=0$, since then
equation
\re{1.7} will be solvable in an interval $(-a,0]$, where $a$ is determined by the
requirement
\begin{equation}
\lim_{\tau\ra -a}F(\f)=0.
\end{equation}

Hence, if $F(\f)$ never vanishes, the solution of \re{1.7} will exist in $(-\un,0]$.
Moreover, in $\tau=-\un$ there cannot be a singularity, a big bang, since this would
require that the mean curvature of the coordinate slices tend to $-\un$. But this
impossible, since $H$ never changes sign, there exist no maximal hypersurfaces in
$N$.

\cvm
To give an explicit example set $\s=\mu=0$ and assume $\tilde\ka=0,-1$. Then
equation \re{1.5} has the form
\begin{equation}
\dot\f^2=-\tilde\ka\tilde\ga^2\f^2+\tfrac{2\ka\tilde\ga^2}{n(n-1)}\rho_0.
\end{equation}
If $\tilde\ka=-1$, we deduce
\begin{equation}
\f=\lam \sinh (c\tau),\q\lam<0, \q c>0,
\end{equation}
and if $\tilde\ka=0$, then
\begin{equation}
\dot\f=-c^2,
\end{equation}
hence
\begin{equation}
\f=-c^2\tau,
\end{equation}
i.e.,
\begin{equation}
e^f=(-c^2\tau)^{\tilde\ga^{-1}}.
\end{equation}

\section{A counter example}\las{3}

We shall show that, even in the case of Robertson-Walker spaces, the transition flow
is in general only of class $C^3$, by constructing a counter example.

\bt
Let $\om=\om_0$ be such that
\begin{equation}
\tilde\ga=\tfrac12(n+\om-2)\ge 2,
\end{equation}
and assume $\s\ne0$. Then the Friedmann equation \re{0.12} has a solution in the
interval $(-a,0)$ such that corresponding spacetime is an ARW space.  The transition
flow $y=y(s)$, however, is only of class $C^3$. If
$\tilde\ga=2$, then $y$ is of class $C^{3,1}$, but, if $\tilde\ga>2$, then
\begin{equation}
\lim_{s\ua 0}\abs{\frac{d^4\h}{(ds)^4}}=\un,
\end{equation}
where $\h=\h(s)$ is defined as in \re{2.18}.
\et

\bp
Due to \rl{1.1} the Friedman equation is solvable and the resulting spacetime is an
ARW space.

Notice that
\begin{equation}
s=
\begin{cases}
-ce^{\tilde\ga f},& s<0,\\
\hp{-}ce^{\tilde\ga f},&s>0,
\end{cases}
\end{equation}
and hence we conclude
\begin{equation}
\h'(s)=\f^{-1},
\end{equation}
\cf equation \re{2.20}, where $\f^2$ can be expressed as
\begin{equation}
\f^2=-\tilde\ka c_1s^2+c_2\rho_0+c_3\s(s^2)^{1+\tilde\ga^{-1}}
\end{equation}
with positive constants $c_i$.

The proof of the theorem can now be completed by elementary calculations.
\ep

\nocite{cg:imcf,cg:indiana,khoury:cyclic,steinhardt:cyclic}


\bibliographystyle{amsplain}
\providecommand{\bysame}{\leavevmode\hbox to3em{\hrulefill}\thinspace}
\providecommand{\MR}{\relax\ifhmode\unskip\space\fi MR }
\providecommand{\MRhref}[2]{%
  \href{http://www.ams.org/mathscinet-getitem?mr=#1}{#2}
}
\providecommand{\href}[2]{#2}



\end{document}